\documentclass[11pt,preprint]{aastex}
\usepackage{graphicx}
\usepackage{natbib}
\bibpunct{(}{)}{;}{a}{}{,} 

\def\eg{{\em e.g.}}

\def\mev{\,{\rm MeV}}

\def\ie{{\em i.e.}}

\def\mz{{\rm Meissner }}

\providecommand{\dfrac}[2]{{\displaystyle\frac{#1}{#2}}}

\shorttitle{Application to Soft Gamma-Ray Repeaters}
\shortauthors{Ouyed et al.}

\begin{document}

\title{3-Dimensional Simulations of the Reorganization of a
Quark Star's Magnetic Field as Induced by the Meissner Effect}

\author{Rachid Ouyed$^{1,2}$, Brian Niebergal$^{1}$, Wolfgang Dobler$^{1}$ and Denis Leahy$^{1}$ }

\affil{$^{1}$Department of Physics and Astronomy, University of Calgary, 
2500 University Drive NW, Calgary, Alberta, T2N 1N4 Canada}
\affil{$^{2}$Canadian Institute for Theoretical Astrophysics, University of Toronto, Toronto, Ontario, Canada }

\email{ouyed@phas.ucalgary.ca}

\begin{abstract}
In a previous paper (Ouyed et al. 2004) we presented a 
new model for soft gamma-ray repeaters (SGR), based on the onset of 
colour superconductivity in quark stars.  In this model, the bursts 
result from the reorganization of the exterior magnetic 
field following the formation of  
vortices that confine the internal magnetic field
(the \mz effect).   Here we extend the model by presenting full 
3-dimensional simulations of the evolution of the inclined exterior magnetic 
field immediately following vortex formation.
The simulations capture the violent reconnection events in the 
entangled surface magnetic field as it evolves into a smooth, 
more stable, configuration which consists of a dipole field 
aligned with the star's rotation axis.  
The total magnetic energy dissipated in this process is found 
to be of the order of $10^{44}\,\mathrm{erg}$ 
and, if it is emitted as synchrotron radiation, peaks typically at
$280\,\mathrm{keV}$.
The intensity decays temporally in a way resembling SGRs and AXPs
(anomalous X-ray pulsars), with a
tail lasting from a few to a few hundred times the rotation period of the
star, depending on the initial inclination between the rotation and
dipole axis.
One of the obvious consequences of our model's final state (aligned rotator)
is the suppression of radio-emission in SGRs and AXPs following their
bursting era.
We suggest that magnetar-like magnetic field strength alone cannot be responsible for the 
properties of SGRs and AXPs, while a quark star entering the ``Meissner phase'' is 
compatible with the observational facts.  We compare our model to observations and 
highlight our predictions.
\end{abstract}
\keywords{gamma rays: bursts --- X-rays: stars --- stars: magnetic 
fields --- stars: neutron --- stars: quark star}


\section{Introduction}

Soft $\gamma$-ray repeaters (SGRs) are sources of recurrent, 
short ($t \sim 0.1\,\mathrm{s}$), intense ($L \sim 10^3\mbox{--}10^4 L_{\rm Edd}$) 
bursts of $\gamma$-ray emission with a soft energy spectrum.
The normal pattern of SGR activity are intense activity periods 
which can last weeks or months, separated by quiescent phases 
lasting years or decades. The five known SGRs are located in 
our Galaxy or, in the case of SGR 0526-66, in the Large 
Magellanic Cloud.  The two most intense SGR bursts ever recorded 
were the 5 March 1979 giant flare of SGR 0526-66 
(Mazets et al.\ 1979) and the similar 27 August 1998 giant 
flare of SGR 1900+14.  The peak luminosities of these events 
($\sim 10^6\mbox{--}10^7 L_{\rm Edd}$) exceeded the peak luminosities 
of ``normal'' SGR bursts by a factor $>10^3$.
Several SGRs have been found to be X-ray pulsars with an 
unusually high spin-down rate of $\dot{P} / P \sim 10^{-10}$~s$^{-1}$, 
usually attributed to magnetic braking caused by a super-strong 
magnetic field $B > 10^{14}$G, which leads to the interpretation that SGRs are 
magnetars (Golenetskij et al.\ 1979; Duncan \& Thompson 1992, 
Kouveliotou et al.\ 1998, Kouveliotou et al.\ 1999). 
In the magnetar model, the magnetic field is the likely provider 
of the burst energy.  A common scenario assumes that magnetic stresses
create a quake in the crust of the
neutron star, which then ejects hot plasma Alfv{\'e}n waves through
its rigid magnetosphere (Thompson \& Duncan 1995; 1996). 
The magnetic field of such a star would have grown to 
magnetar-scale strengths because of strong convection during 
the collapse of the proto-neutron star core 
(Duncan \& Thompson 1992; Thompson \& Duncan 1993).

\subsection{Open issues in the magnetar model of SGRs}
\label{sec:open_issues}
In the magnetar model of SGRs, which is also that of 
Anomalous X-ray Pulsars (AXPs), the X-rays are ultimately 
powered by an internally decaying very strong magnetic field.  
However there are still a few open questions which in our opinion leave
room for new models to be explored:
   
\begin{itemize}
\item  Despite numerous attempts, no magnetars have been detected 
at radio frequencies\footnote{Detecting radio pulsations may be difficult,
given the small polar caps implied by the long spin periods.}
(Kriss et al. 1985; Coe et al. 1994; 
Lorimer et al. 1998; Gaensler et al. 2001).
 It has been suggested that QED processes at high $B$, such as photon 
splitting, may preclude the electron/positron cascades necessary 
to produce radio emission (Baring\&Harding 2001), or that   
 pair production ceases above some critical magnetic field (Zhang \& Harding 2000).
 These ideas or alternatives remain to be confirmed and are still debatable. 
\item One might expect high-$B$ radio pulsars to be more X-ray bright 
than low-$B$ sources, and to possibly exhibit AXP-like burst emission. 
However X-ray observations of 5 high-$B$ radio pulsars reveal luminosities 
much smaller (by a few orders of magnitudes) than those of AXPs 
(Pivovaroff et al. 2000; Gonzalez \& Safi-Harb 2003; McLaughlin et al. 2003; 
Gonzalez et al. 2004; Kaspi \& McLaughlin 2005). 
This has lead to suggestions that high-$B$ radio pulsars may one day 
emit transient AXP-like emission, and conversely that the
transient AXPs might eventually exhibit radio pulsations 
(Kaspi \& McLaughlin 2005) -- a notion yet to be confirmed.
\item  Hints of  massive ($> 30\mbox{--}40\,M_{\odot}$) progenitors
 associated with AXPs and SGRs by recent observations 
(Gaensler et al. 2005) has led to the suggestion that pulsars
 and SGRs differ in their progenitor masses. It is also suggested that
 massive progenitors could lead to neutron stars with millisecond periods (Heger et al. 2004) which
 would comply with the magnetar model for SGRs (Duncan \& Thompson 1992).
This, however,
leaves open the question of why high-$B$ pulsars,  formed from less massive progenitors
 (presumably with periods $> 15$ ms), possess magnetar-like field strengths.
\item All SGRs and AXPs known to date have spin periods between 6 and 12 seconds
(Kaspi 2004).  This clustering in period remains to be explained by
the magnetar model.
\end{itemize}

Here we explore an alternative model  
first presented by Ouyed et al. (2004) where one
assumes that AXPs and SGRs are quark stars, rather than magnetars. 
While quark stars have yet to be found in nature, formation
scenarios have been suggested in the literature
(see Sect.~8.4 in  Ouyed et al. 2004 and references therein).  
The qualitative idea is that the core of a neutron star reaches 
deconfinement densities, eventually leading to the conversion of the entire
 star to a quark star. 
In principle, the transition from hadronic matter into quark matter in the
core of a neutron star can 
happen immediately during or after the supernova explosion, but also 
much later than that.  Such a transition could occur in a smooth 
stable manner (e.g. Bombaci \& Datta 2000 and references therein) or in 
an explosive manner termed the ``Quark-Nova''
(Ouyed et al. 2002; Ker\"anen, Ouyed, \& Jaikumar 2005).  In these
formation scenarios, if the quark star  
 is born shortly after the supernova, 
the emitted energy/radiation would be absorbed by the still expanding supernova ejecta, 
and there would be no detectable signature of the transition. 
The hadron-quark transition may also happen much later than the supernova 
explosion and could be induced, for example, by accretion if the neutron 
star is a member of a binary, or via spin-down in the case of isolated neutron stars.
The new idea of this model, first presented in Ouyed et al. (2004), is
that the quark star
enters a superconductive phase, and subsequently experiences
a ``Meissner phase'' that triggers the reorganization of the
star's magnetic field. Before going into more details we first 
describe properties of quark matter and the concept of color superconductivity.

\subsection{Superconductivity and Meissner effect in quark stars}

The discovery of asymptotic freedom, leading to the formulation of
quantum chromodynamics (QCD) as the theory of strong interactions,
was soon followed by the suggestion that matter at sufficiently
high densities consists of a deconfined phase of
quarks \citep{col:1975}. Only shortly afterwards it was pointed
out \citep{bar:1977} that the true ground state of cold dense quark
matter exhibits color superconductivity (CSC), characterized by
diquark condensation with an estimated energy gap $\Delta \simeq 1\,\mev$
between the highest occupied and the unperturbed
quark state at the Fermi surface. Since this magnitude of the gap is 
rather small for phenomenological applications, CSC subsequently 
received little attention. The situation changed when
reinvestigations \citep{alf:1998,rapp:1998}, using nonperturbative
forces (\eg, instanton-induced), showed that the gap can be
substantially larger, $\Delta\simeq100\,\mev$ for moderate
quark chemical potentials, $\mu_q\simeq350\,\mev$. Similarly
large values are obtained from estimates based on perturbative
calculations at asymptotically high densities \citep{pis:1999,son:1999}.
Thus, from the practical point of view, the existence of color 
superconductivity in compact stars has (re-)emerged as an exciting 
possibility.

The detailed properties of CSC matter relevant to astrophysical
applications depend on the interplay of the quark chemical potential,
the $q$-$q$ interaction strength, and the bare masses of the (light)
quarks $u$, $d$ and $s$. In particular, for $\mu_q$ below the
(constituent) strange quark mass, only $u$ and $d$ quarks are
subject to Cooper pairing. The corresponding phase is known as
2-flavor CSC (2SC). In the idealized case where the quark chemical
potential is much larger than the strange quark mass ($m_s$), the
latter becomes negligible and all three flavors exhibit likewise
pairing. The preferred symmetry (breaking) pattern in this phase
corresponds to the so-called color-flavor locking
\citep[CFL;][]{alf:1999}, since the underlying diquark condensate is
invariant only under simultaneous color and flavor
transformations. In the present work, we will focus on the
CFL phase (for a recent review and a more exhaustive list of
references, cf.~\citealt{sch:2003}).
Associated with CSC is the critical temperature, 
$k_{\rm B}T_{\rm c}\sim 0.5\,\Delta$, above which pairing is washed out. 

One of the most interesting properties of an ordinary
superconductor is the Meissner effect, i.e., the
expulsion of magnetic flux from the superconductor \citep{meissner}. 
In the CFL phase, the gauge bosons connected with the broken
generators obtain masses, which indicates the Meissner
screening effect \citep{rischke00a,rischke00b}.
This is the heart of our model which we describe next.

\subsection{Our model}

\begin{figure}[ht]
\plotone{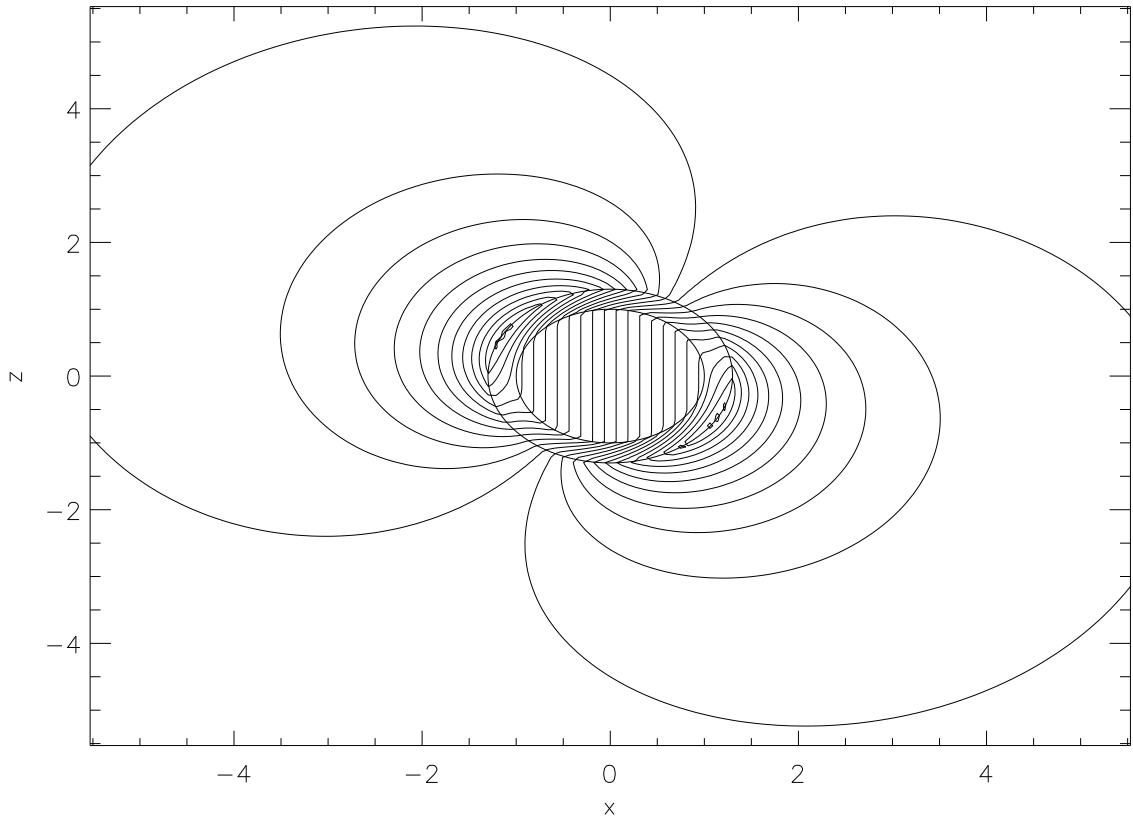}
\caption{``Mock magnetic field lines'' at $t=0$ in the $xz$-plane.
The rotation axis is in the z direction inclined by an angle $\theta$ with respect to the dipole axis.
Initial state after the stars interior converts to the CFL phase thus causing
the magnetic field to be confined inside the vortices. The inner circle represents 
the surface of the star, and between that and the outer circle is the region 
where surface currents cause the outer dipole field to adjust to the
vortex-aligned inner field.
Shown are isolines of the vector potential component $A_y$, which for our
setup trace most features of magnetic field lines.
}\label{fig:init_state}
\end{figure}

Assume a quark star is born with a temperature $T > T_{\rm c}$
and enters the CFL phase as it cools by neutrino emission
(Ker\"anen, Ouyed \& Jaikumar 2005).
The CFL front quickly expands to the entire star followed
by the formation of rotationally induced vortices,
analogous to rotating superfluid $^3$He (the vortex lines 
are parallel to the rotation axis; Tilley\&Tilley 1990).  
Via the \mz effect, the magnetic field is partially screened 
from the regions outside the vortex cores. 
The system now consists of alternating 
regions of superconducting material with a screened magnetic 
field, and the vortices where most of the magnetic field resides. 
As discussed in Ouyed et al. (2004), this has interesting consequences
on how the surface magnetic field can adjust
to the interior field which is pinned to the vortices.
Figure~\ref{fig:init_state} shows the starting point for this
reorganization of the magnetic field, just after the Meissner effect has
fully aligned the magnetic flux inside the star with the rotation axis.
Within a transition region, the magnetic field switches from the vertical
interior field to that of an inclined dipole outside.
A conservative assumption (from the energetics point of view) is that the
field in the transition region is a potential field (minimum-energy
configuration).

In order to capture the complex
non-linear dynamics of the system 
we need the help of 3-D simulations. These simulations will allow us
to follow the evolution and re-organization of the magnetic field 
induced by the vortices.

\begin{figure}[ht]
\plotone{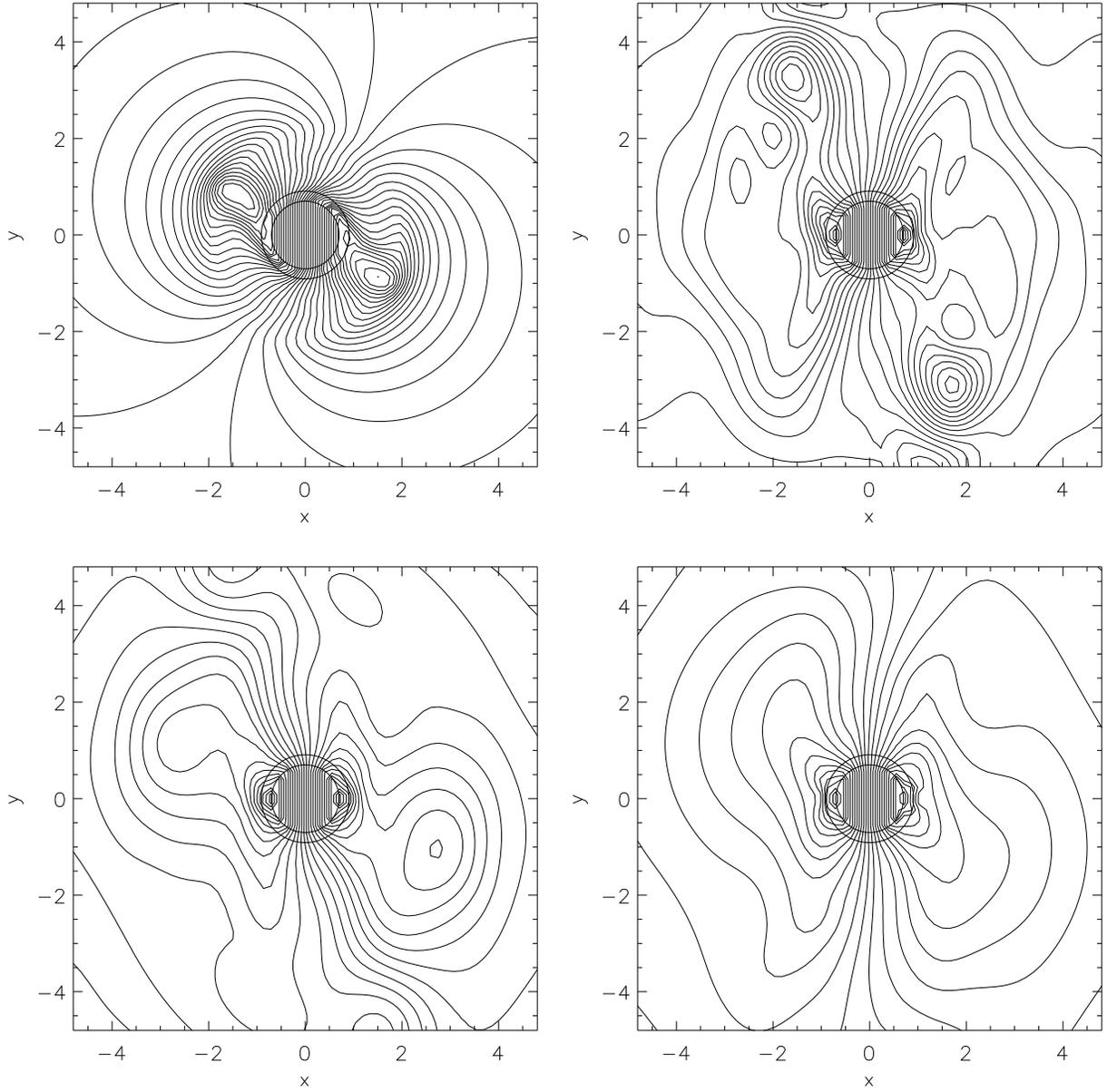}
\caption{``Mock magnetic field lines'' in the $xz$-plane (rotation axis is
  along z)
at $t=1$, $25$, $100$, $300$ going left to right, top to bottom. Note the 
Alfv\'en front at $t=1$, which is accompanied by reconnection events. 
Reconnection events become more apparent with time as the surface magnetic
field re-adjusts itself.  In the last frame we see an aligned dipole which
would mark the end of radio activity.}
\label{fig:xz_plane}
\end{figure}
\begin{figure}[ht]
\plotone{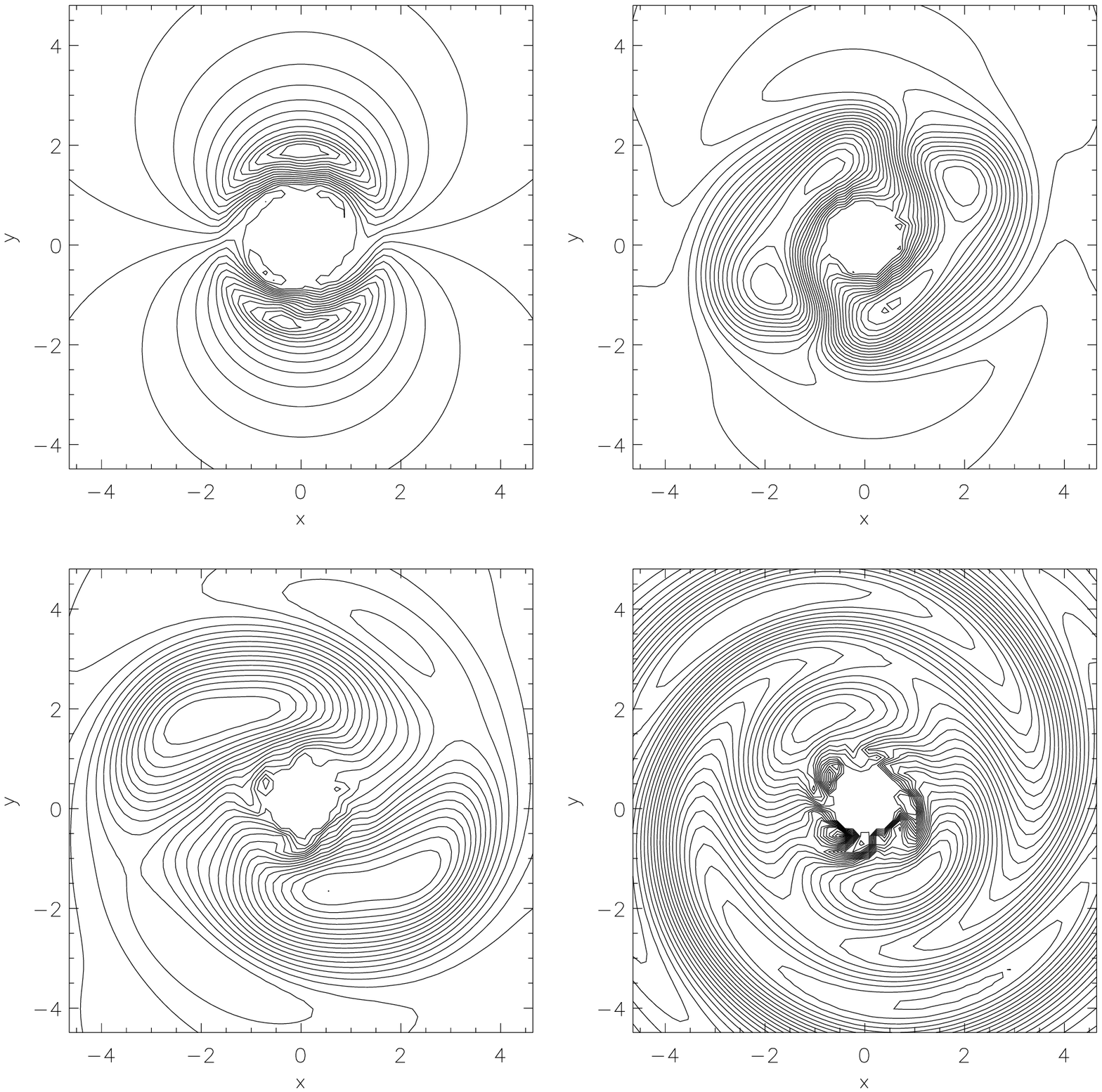}
\caption{\label{fig:xy_plane}``Mock magnetic field lines'' in the $xy$-plane (down the
  rotation axis) at $t=1$, $25$, $100$, $300$ going left to right, top to bottom.}
\end{figure}

The paper is presented as follows:
In \S~\ref{sec:setup} we describe and analyze the basic setup of our
simulations.
We calculate the 
synchrotron light curves that result from the simulations wherein
we find remarkable similarities to SGR light curves. 
Mechanisms for the observed subsequent bursts and the quiescent phase 
in SGRs are then briefly discussed in terms of our model
in \S~\ref{sec:vortices}.
In \S~\ref{sec:predict} we discuss the model predictions
and how it can account, at least in its
current stage, for the points listed in \S~\ref{sec:open_issues}.
Here we suggest a list of observations that could test our model. 
We conclude in \S~\ref{sec:conclusion}.

\section{Simulations}
\label{sec:setup}
Two fundamental parameters of our model are the inclination angle $\theta$
of the external dipole field relative to the rotation axis,
and $\beta$, the initial ratio of gas to magnetic pressure at the surface.
The simulations start with the interior magnetic field
confined in vertical vortices for $r<R_{\rm QS}$, but with a still
unperturbed inclined dipole field for $r > 1.3\,R_{\rm QS}$.
The transition region $R_{\rm QS}<r<1.3\,R_{\rm QS}$ is filled by a
potential field and is bounded by two current layers
(see Fig.~\ref{fig:init_state}).
The whole configuration was chosen by minimizing the total magnetic energy
(see Appendix~\ref{sec:FieldConf}).

We solve the following set of non-ideal magnetohydrodynamical (MHD)
equations, using the \textsc{Pencil-Code} \citep[see \eg][]{dob:convstar}\footnote{
  http://www.nordita.dk/software/pencil-code --
  The Pencil Code is a high-order finite-difference code for solving the
  compressible hydromagnetic equations.
}.
\begin{eqnarray}
\label{Eq-continuity}
\frac{D\ln{\varrho}}{Dt} &=& -\nabla\cdot\vec{u} \\
\frac{D\vec{u}}{Dt} &=& -c_s^2\nabla\left(\frac{s}{c_p} +
  \ln{\varrho}\right) - \nabla\Phi_{\rm grav} +
  \frac{\vec{j}\times\vec{B}}{\varrho} \\
  \nonumber &+& \nu\left(\nabla^2\vec{u} +
  \frac{1}{3}\nabla\nabla\cdot\vec{u} +
  2\mathsf{S}\cdot\nabla\ln{\varrho}\right) \\
\frac{\partial \vec{A}}{\partial t}
  &=& \vec{u}\times\vec{B} - \eta\mu_0\vec{j}\\
  \label{Eq-entropy}
  \varrho T\frac{Ds}{Dt} &=& \nabla\cdot\left(K\nabla T\right) +
  \eta\mu_0\vec{j}^2 + 2\varrho\nu\mathsf{S}^2 \; .
\end{eqnarray}
Here $\varrho$, $\vec{u}$, $\vec{A}$, and $s$ are density, velocity,
magnetic vector potential, and specific entropy, 
respectively.  Parameters and functions kept constant were 
$c_p$, $\Phi_{\rm grav}$, $\nu$ and $\eta$ corresponding to 
specific heat, gravity potential, 
kinematic viscosity and magnetic diffusivity respectively.
The remaining variables $c_s$, $\vec{j}$, $\vec{B}$, $\mathsf{S}$ and
$T$ represent the sound speed, electric current density, magnetic flux
density, traceless rate-of-strain tensor, and temperature, respectively.

The length scale is in units of the radius of the quark star, $R_{\rm QS}$, 
density in units of the star's surface density, $\rho_0$, and time is in
units of the spin-period, $1/\Omega$. This implies that velocities are
in units of $R_{\rm QS}\Omega$, while the magnetic field is in units of
$\sqrt{\rho_0} R_{\rm QS}\Omega$. The strength of the (dipole) magnetic
field at the surface can be estimated for a given $\beta_0$ as, 
$B_0^2 = 8\pi P_0/\beta_0$, where $P_0$ is the pressure at the surface of the star.
Using hydrostatic balance and the perfect gas law this becomes,
\begin{eqnarray} 
B_0 &\sim& \frac{5\times 10^{13}~{\rm G}}{\sqrt{\beta_0}}\times \\
\nonumber
&&\left(\frac{\rho_0}{10^6~{\rm g/cm^3}}\right)^{1/2}\left(\frac{10~{\rm km}}{R_{\rm QS}}\right)^{1/2}\left(\frac{M_{\rm QS}}{M_{\odot}}\right)^{1/2} .
\end{eqnarray}
In the equation above, $\rho_0$ is 
the average density of the gas close to the surface
 of the star. This corona is supplied
 by  fall-back material following the formation of the QS (Ker\"anen, Ouyed, \& Jaikumar 2005)
 similar to what has been suggested in the supernova case \citep{Chev89}.  
The density of the  fallback matter,  representative
 of the  crust material of the parent neutron star, is estimated
 to be of the order of $10^6~{\rm g~cm}^{-3}$. 

Note that our MHD equations (\ref{Eq-continuity})--(\ref{Eq-entropy}) are
non-relativistic.
While near the surface of the quark star some aspects of the physics will
be considerably changed by relativistic effects, we expect that the
overall dynamics will not be vastly different in a more realistic
calculation.

All figures shown here are for resolution $128^3$.
We have run simulations at higher resolution, but found that they differ
little as far as energetics and evolution are concerned.

\subsection{Evolution and reorganization of the surface magnetic field}
\label{sec:results}
Figures~\ref{fig:xz_plane}--\ref{fig:xy_plane} show the evolution of the exterior
magnetic field as it adjusts to the vortex-confined field\footnote{See the`` Animations''
link at http://www.capca.ucalgary.ca for movies of the simulations.}. 

The complicated structure of the surface magnetic field is clearly 
seen in the animations driven
by the frequent magnetic reconnections as the surface field tries
to align itself with the interior one (rotation axis).
These random reconnection events
would bear many similarities to the 
initial events (\ie~those near $t=0$), but we expect them
to be less energetic as the magnetic field slowly decays and weakens. 
Eventually, the magnetic field evolves into a stable configuration 
(see Fig.~\ref{fig:xz_plane}) after which the star enters a quiescent phase.

The restructuring of the field in the transition region leads to an
approximately
spherical Alfv{\'e}n wave traveling outwards (see Fig.~\ref{fig:xz_plane} for $t=1$) and
is more prominent in simulations with a stronger magnetic field (\ie~smaller $\beta$).
As the wave travels outwards it amplifies the magnetic field in certain regions causing 
them to undergo reconnection, which both distorts the wave and eventually damps it out.
Furthermore, the regions that underwent reconnection appear to show
slow oscillatory motions between the reconnection site and the surrounding
gas (``breathing'').
This can be seen in the series of diminishing pulses in
Fig.~\ref{fig:intensity_time} and the frequency of the pulses remains nearly constant
(see Fig.~\ref{fig:fft}).  We note that these pulses appear more prominently in simulations
with lower $\beta$ and do not arise in simulations with $\beta > 1$. 

\subsection{Energetics and emission}
\begin{figure}[ht]
\plotone{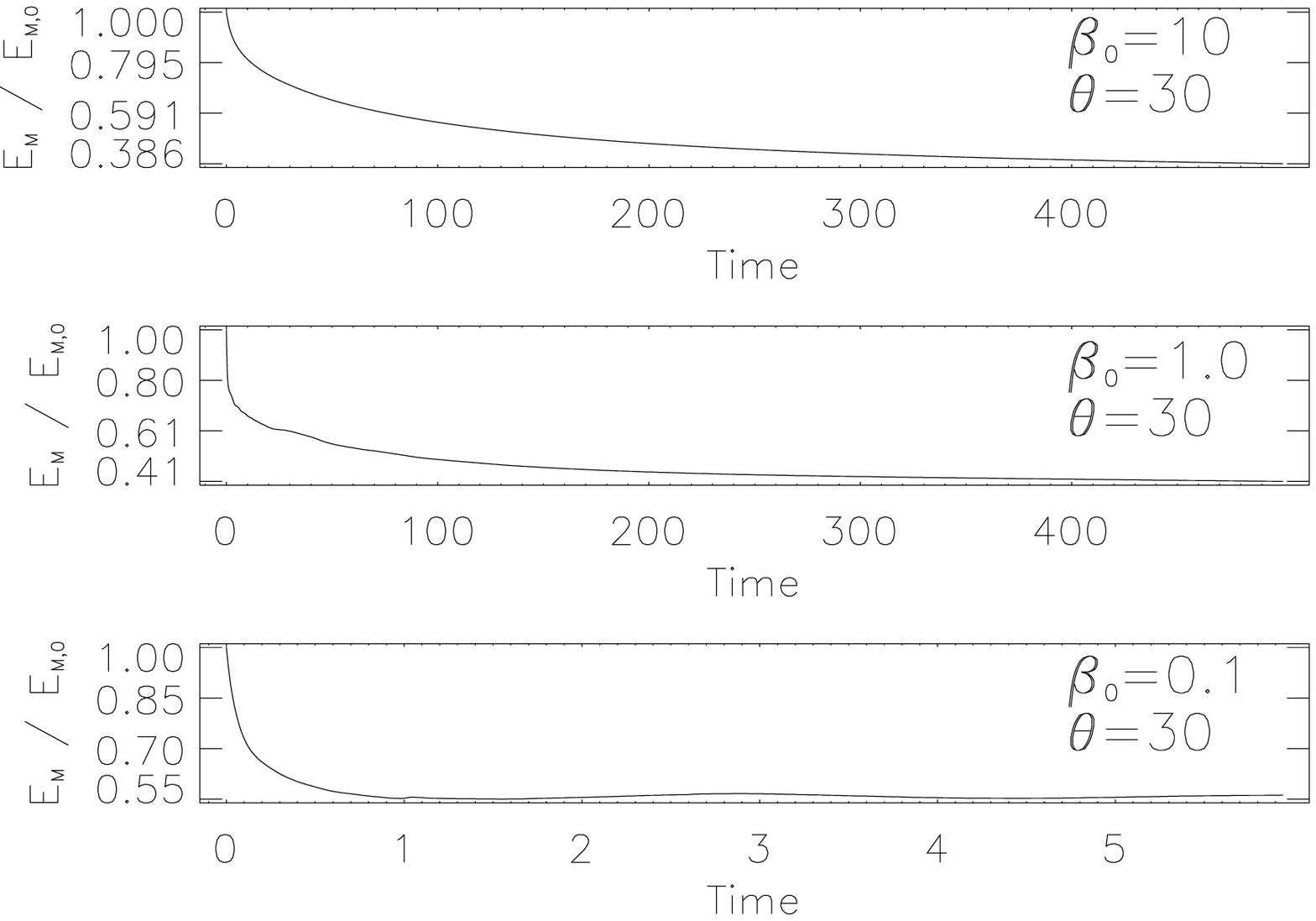}
\caption{\label{fig:em_time}Magnetic energy lost over time. The change 
in magnetic energy ($\alpha$ in Eq.~\ref{equ:EM2}) can be estimated 
from these plots to be between $0.4-0.6$, implying that energies
of order $10^{44}~{\rm erg}$ are released during the event. }
\end{figure}

\begin{figure}[ht]
\plotone{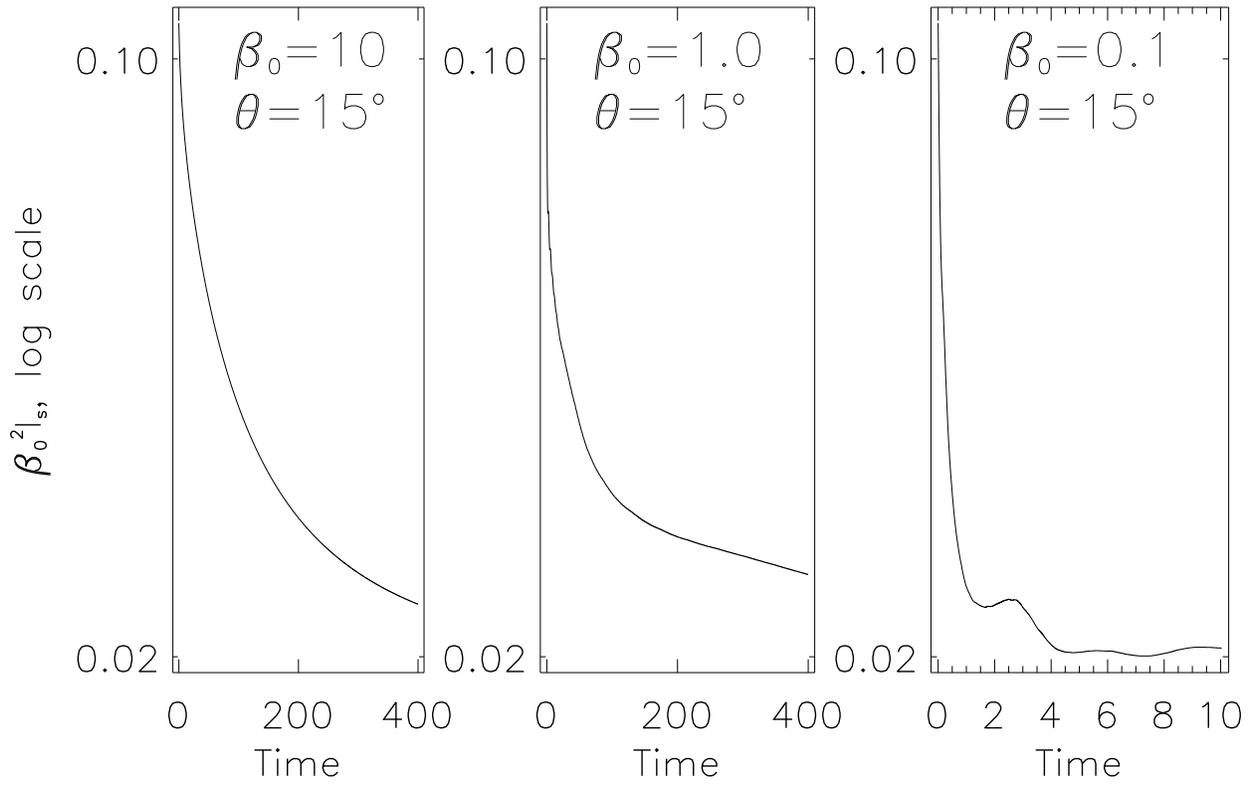}
\caption{\label{fig:intensity_time}Total intensity emitted ($\beta_0^2 I_{\rm s}$) versus time 
for $\theta=15^\circ$ and $\beta = 10, 1, 0.1$.
We note that the higher $\beta$, the slower the re-adjustment and flattening of the tail.
The decrease in variability with increasing $\beta$ can be explained by a decrease in
amplitude of magnetic ``breathing'' modes.}
\end{figure}

The magnetic energy released in the organization is shown in
Fig.~\ref{fig:intensity_time} and can be cast 
into a simple equation,
\begin{equation} \label{equ:EM2}
E_{\rm M} \sim 10^{44}\, {\rm erg} \left(\frac{\alpha}{0.5}\right)^2\left(\frac{B}
{5\times 10^{13}\, {\rm G}}\right)^{2}\left(\frac{R_{\rm QS}}{10\, {\rm km}}\right)^{3}\ ,
\end{equation}
where $\alpha$ is the fraction of the surface magnetic
field that decayed via reconnection events. From Fig.~\ref{fig:em_time}
we can infer the decay in magnetic energy to be roughly $\alpha=$ $0.4$ - $0.6$.

Furthermore, if we assume the electrons to be co-rotating
($\gamma_e \sim 1$), then the intensity emitted by synchrotron 
processes from the simulated region goes as 
$I_{\rm s} \propto n_e B^2$. Also, assuming peak emission is 
at $\nu_{\rm p} = 0.29\nu_c$, where $\nu_c$ is the critical frequency 
(see Sects.~6.2 and 6.4 in Rybicki \& Lightman 1979), 
we find this emission to be in the X-ray band,
\begin{eqnarray}
h\nu_{\rm p} &\sim& 335~{\rm keV}\left(\frac{B}{5\times 10^{13}{\rm G}}\right) \times \\
\nonumber
&&\left[1-0.29\left(\frac{M_{\rm QS}}{M_{\odot}}\right)\left(\frac{10~{\rm km}}{R_{\rm QS}}\right)\right]^{1/2} ,
\end{eqnarray}
where gravitational redshift is included in the last term.
This intensity decays temporally in a 
way closely resembling SGR bursts (see Fig.~\ref{fig:intensity_time}). 
The decay profiles of our simulated bursts depend on the initial plasma
$\beta$ at the surface of the star and on the dipole inclination $\theta$. 
Varying these parameters in our model allows us to fit the exponential
decay shape and duration of observed bursts.

\begin{figure}[ht]
\plotone{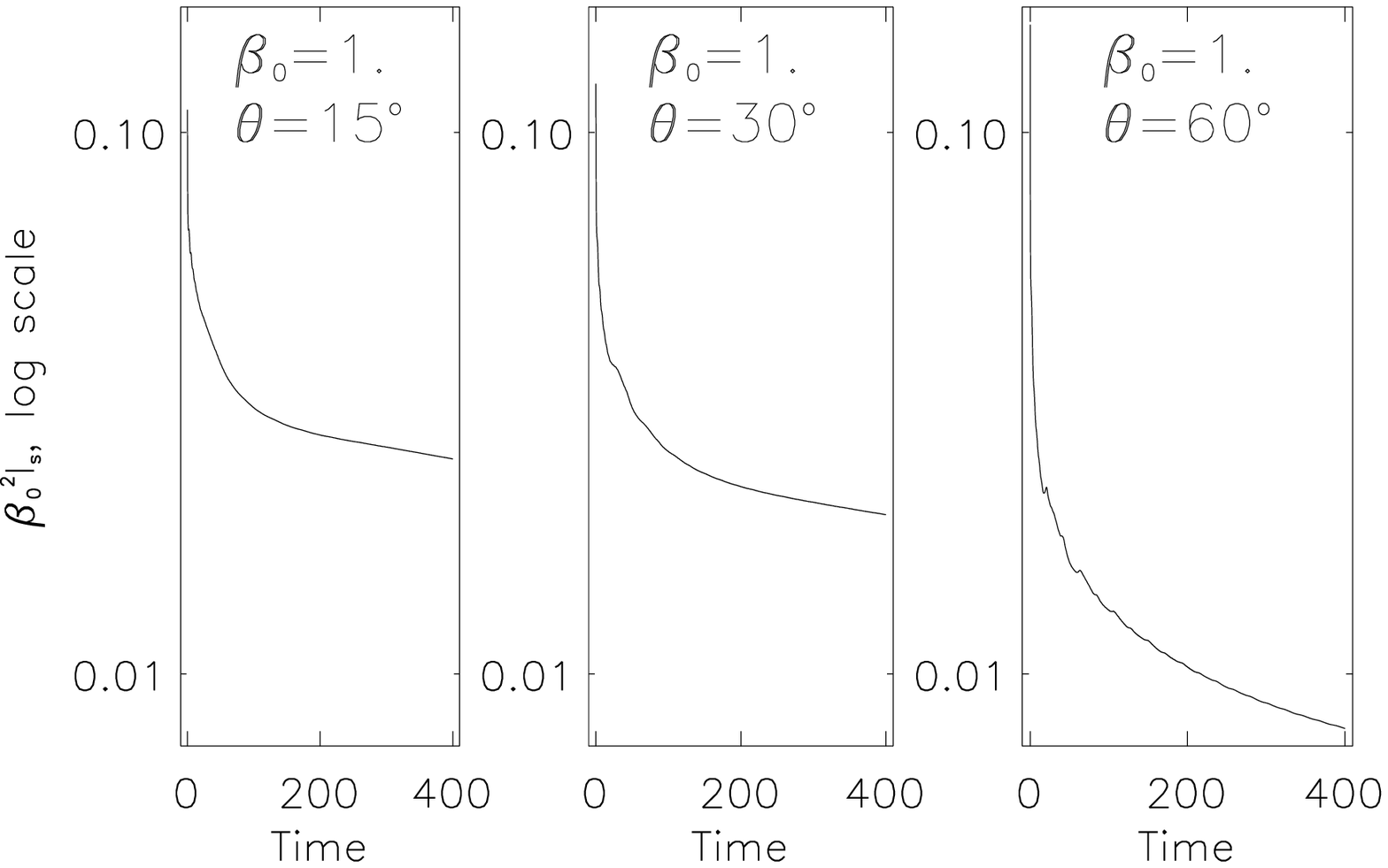}
\caption{\label{fig:beta_theta}Total intensity emitted ($\beta_0^2 I_{\rm s}$) versus time
for $\beta=0.3$ and $\theta=15^\circ, 30^\circ, 60^\circ$ (angle between
rotation and initial magnetic dipole axis).
We note that the higher the inclination 
angle, the longer it takes for the outer magnetic field to align itself.
This can be
seen in the figures where the tail takes longer to flatten for higher angles.
Furthermore, the high inclination run shows oscillations which can be linked to
stronger reconnection events.}
\end{figure}

\subsection{Periodicity in emission}\label{sec:periodicity}
\begin{figure}[ht]
\plotone{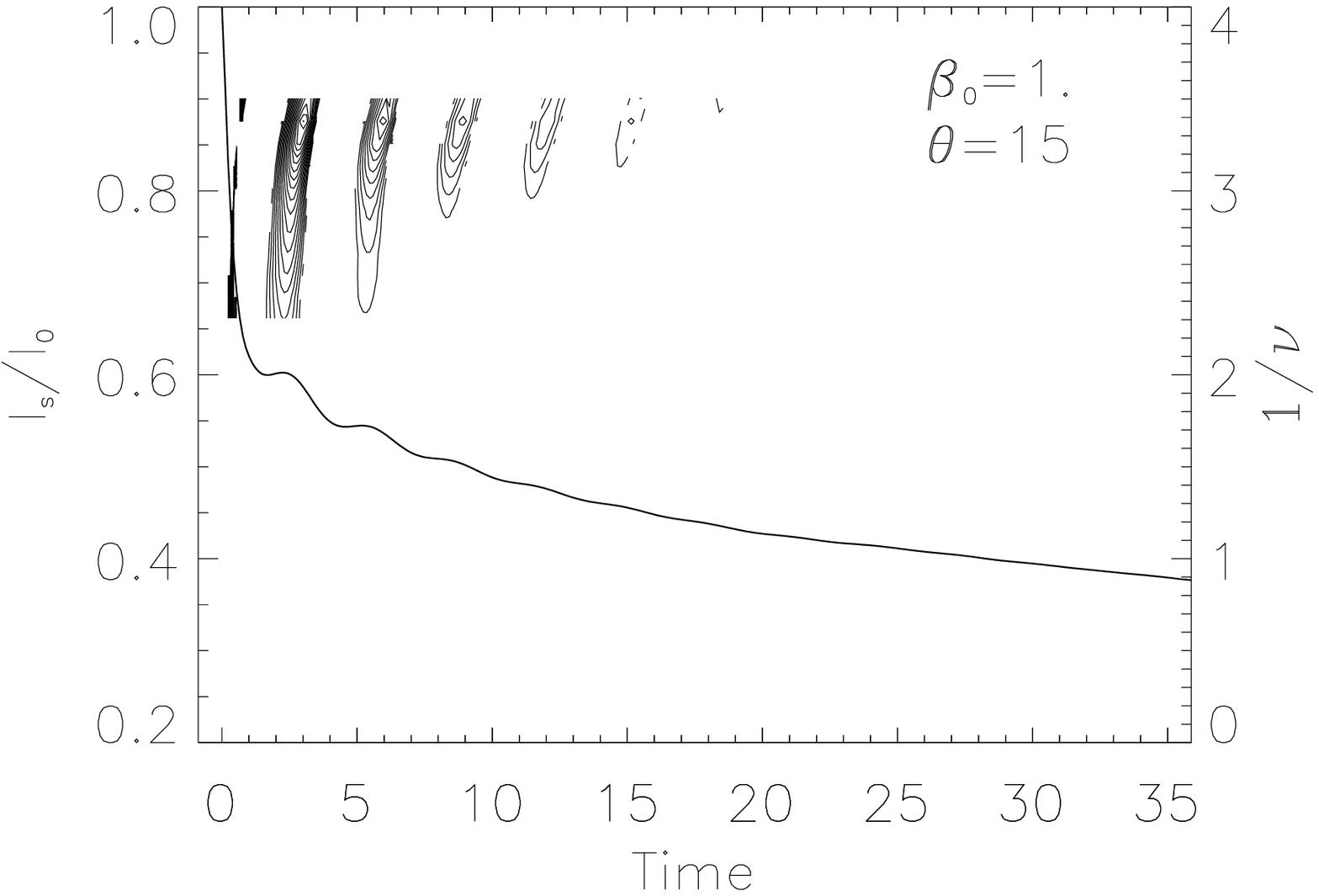}
\caption{\label{fig:fft}Time evolution of the Fourier-power spectrum 
(contours, right axis) for $\beta=1$ and $\theta=15^\circ$ along 
with the normalized intensity (solid line, left axis). 
The sub-pulses in intensity are shown here to be modulated
at $\sim 3.25$ times the spin-period.
In an observer's frame, this modulation should
be further modulated by the spin period.  }
\end{figure}
The oscillations shown in Fig.~\ref{fig:fft}, can be interpreted as magnetic
``breathing'' modes.  In this situation, regions where magnetic reconnection 
occurs cause a disparity in the magnetic pressure between neighbouring regions. 
This low pressure region will draw plasma in from the surrounding regions
causing either more reconnection or enhanced density where, in either case,
there will be oscillations in the emission (\ie~$n_e B^2$).

Furthermore, since our simulation is performed in the 
co-rotating frame, our model is too simplistic to reproduce
a spin-modulated pattern (e.g. the 8.0 s period observed in the 
March 5 event) superimposed on a smooth exponential decay in the light curves. 
However, as can be seen in Figs.~\ref{fig:xz_plane}--\ref{fig:xy_plane},
there are hotter regions with more magnetic reconnection events than others,
and so spin modulation should occur naturally.
We note that a few of these hot spots appear simultaneously at 
random locations, implying spin-modulation may consist of even smaller sub-pulses,
therefore producing many harmonics in the light curve.
In other words, if our model is a correct representation of SGRs, 
observations could constrain the number of hot spots.

\section{Subsequent bursts and quiescent phase}
\label{sec:vortices}
We note that since the CFL phase transition occurs only once in a given star, 
there is only one giant burst in our model.
However, if we assume some form of heating, then it is possible that the quark star
can revert to a non-CFL state wherein superconductivity is lost, and the 
interior magnetic field lines are no longer constrained to the vortices. 
Then presumably the dynamics that initially caused the magnetic dipole axis to misalign 
with the rotation axis, we speculate, should once again ensue to some extent.
Thus we allow for the possibility of subsequent
misalignment of the magnetic axis that, upon further cooling back below
the deconfinement temperature, can trigger the same process as before to give 
subsequent bursts. One might expect
these bursts strengths to depend on the amount of heating and resulting 
misalignment.  Possible sources of subsequent heating can include accretion
from a companion, impact from the accretion of a small body, or the quark star 
passing through a higher density region
in the ISM.
This leads to the prediction that radio emission should pick up again slightly
before the subsequent bursts, provided there is a favourable line-of-sight. 
The accurate measurement of the temperature in the star in these subsequent 
bursts would now allow observations of the deconfinement
temperature, which has eluded QCD physicists so far.

The quiescent phase is due to vortex expulsion from spin-down and subsequent 
annihilation through magnetic reconnection near the surface.  The number of 
vortices decrease slowly with spin-down leading to continuous, quiescent 
energy release which can last for $10^3$ to $10^4$ years (see Ouyed
et al. 2004 for more details).

\section{Model predictions and observational tests}\label{sec:predict}
In the light of the results presented above we will now discuss 
our model predictions and offer our interpretation of the
open issues listed in \S~\ref{sec:open_issues}:

\begin{itemize}
\item Following reorganization of the outer magnetic field and its
alignment with the rotation axis, our model naturally predicts the
suppression of further radio pulsations.
Our model can thus in principle explain why SGRs and AXPs stop
pulsating following their bursting era. 

\item The high-$B$ (magnetar-strength) radio pulsars that 
show no evidence of enhanced X-ray emission,
can be accounted for in our model if they are just neutron
stars that have not experienced the \mz effect.
As long as the object is a quark star in a superconducting
phase, magnetic field decay and reorganization will take effect
regardless of the field strength.
We thus predict that SGR-like bursts with moderate-strength magnetic field
may be discovered in the future.
               
\item If observations do confirm that the progenitors
of SGRs are very massive stars (Gaensler et al. 2005) this would 
strengthen our model since massive progenitors
are more likely to lead to massive neutron stars, for which
it should be easier to reach deconfinement densities in the
core following accretion or spin-down.

\item The period clustering of observed SGRs and AXPs can be 
explained in our model as the time necessary for the parent
neutron star to cool down sufficiently to experience a 
transition to a quark star.
\end{itemize}       

Other predictions from our model are: 
\begin{itemize}
\item The association of the
SGR source (the quark star) with a parent radio-pulsar.  In other words,
in at least a few cases (if beaming is favorable) a parent radio-pulsar should
be detected in the same location in the sky as the SGR before the bursting activity.

\item If observations show quiescent emission before and after the burst then
this is a subsequent burst in our picture, meaning the SGR will
have been a quark star for some time.  In this case the association with
a parent neutron star will be less obvious.

\item Assuming synchrotron dominated
emission, we should observe a peak at $\sim 280~{\rm keV}\left(B/5\times10^{13}~{\rm G}\right)$
for a solar mass quark star with a $10~{\rm km}$ radius.

\item The total energy in bursts from moderate-strength magnetic 
field ($10^{12}{\rm G}$) quark stars would be much weaker 
according to Eq.~\ref{equ:EM2}. We predict burst energy from these stars 
to be of the order of $10^{40}{\rm erg}$.

\item Finally, we note that during the stars phase transition from hadronic 
to quark matter, its radius changes from $30\%$ - $50\%$ (Ouyed et al. 2002), 
naturally resulting in a significant increase of the magnetic field at the surface.
Therefore, a quark star with a strong magnetic field is not necessarily the result 
of a parent star with a strong magnetic field.
So, if quark stars undergoing the Meissner effect are
indeed the cause of SGRs or AXPs, then stronger field strengths in these objects should be expected.
\end{itemize}

\section{Conclusion}
\label{sec:conclusion}

We presented 3-D simulations of the reorganization of the magnetic field
surrounding a newly born quark star.  The reorganization is a consequence
of the star entering the CFL phase, confining the interior field to vortices,
and leaving the exterior field in its tilted dipole configuration.
In our model the bursting activity in SGRs and AXPs is explained as magnetic reconnection
events occurring while the exterior field aligns itself with the interior one.
One of the obvious consequences of the final state (aligned rotator)
is the suppression of the radio-emission in SGRs and AXPs following their bursting era.
We should emphasize that magnetic fields alone cannot be responsible
for the properties of SGRs and AXPs, but rather a compact star experiencing the 
Meissner effect is required.  A further consequence is that while a low-$B$
quark star entering the ``Meissner phase'' would burst in X-ray, this will not necessarily
be the case for a high-$B$ neutron star.
While quark
stars have not yet been observed in nature, our model seems to account for
many observed features in SGRs and AXPs, thus warranting further investigation.

\begin{acknowledgements}
We thank Ralph Pudritz, Chris Pethick, and Kaya Mori for discussions. 
B.~N. thanks C.I.T.A. for hospitality, 
and Sigma-Xi for its Grant-in-aid of research.
The research of R.~O. is supported by grants from the Natural Science and
Engineering Research Council of Canada (NSERC) and Alberta Ingenuity Fund (AIF). 
\end{acknowledgements}

\appendix

\section{Magnetic energy of a vertical field in a sphere}\label{sec:Etot-Bvert}

Consider a strictly vertical magnetic field
$\vec{B} = B_z(x,y) \hat{\vec{z}}$ in a 
sphere of radius $R$.
\begin{figure}[ht]
  \centering
  \plotone{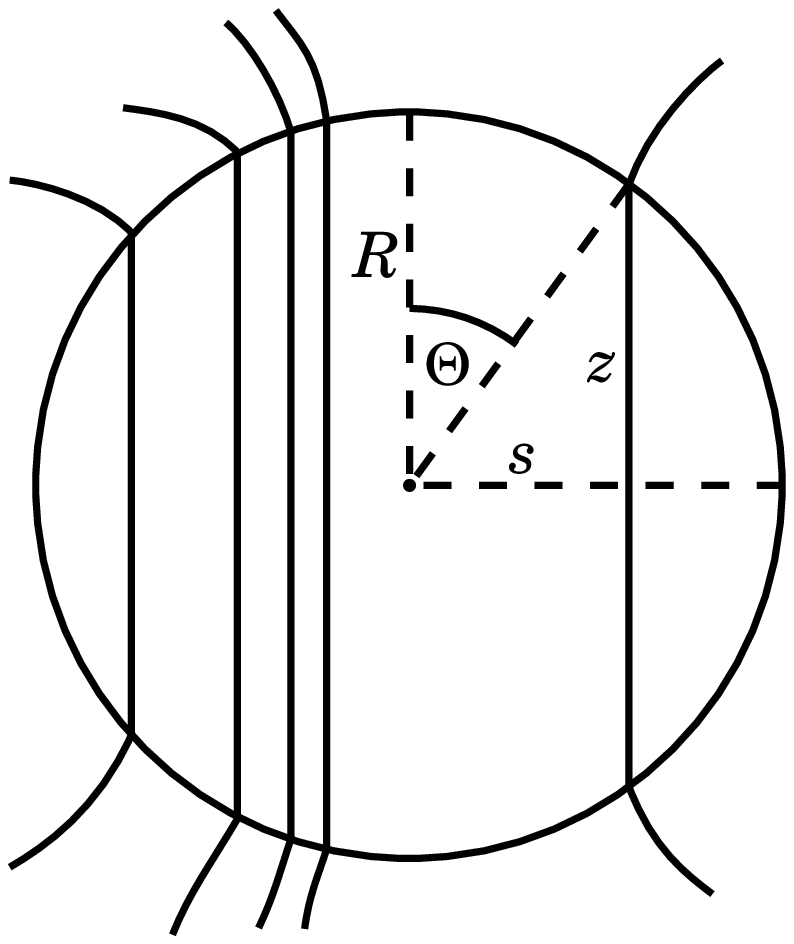}
  \caption{\label{Fig-neutron-star}Sketch of magnetic field inside and outside the sphere.} 
\end{figure}
Since the radial component of $\vec{B}$ must be continuous at the surface,
the field inside the sphere must be
\begin{equation}
    B_z(r,\vartheta,\varphi) = \frac{B_r(R,\Theta,\varphi)}{\cos\Theta} \; ,
\end{equation}
where $\Theta$ is the colatitude at the surface, which is related to
$r,\vartheta$ via
\begin{equation} \label{Eq-r-theta-s}
  r\sin\vartheta = R \sin\Theta = s \; ,
\end{equation}
with $s$ the cylindrical radius.
To calculate the magnetic energy inside the sphere, we integrate
$\vec{B}^2$ over $z$, then transform the $s$ integral in one over
$\Theta$, using Eq.~(\ref{Eq-r-theta-s}).
After a bit of algebra, we find the remarkable expression
\begin{equation} \label{Eq-E-inner}
  E_{\rm mag} = R^3 \int\limits_0^{2\pi} d\varphi\,
          \int\limits_0^{\pi} d\Theta\, \sin\Theta
            \frac{B_r^2(R,\Theta,\varphi)}{2\,\mu_0} \; .
\end{equation}

\section{Initial magnetic field configuration}\label{sec:FieldConf}

The initial magnetic field represents the phase where the Meissner
effect has forced the field inside the neutron star to be strictly
vertical, but this reorganization has not yet had time to affect the
external field.

In order to keep magnetic energy finite, we have to allow for a transition
layer between the vertical internal field and the inclined dipolar
external field.
We are thus looking for a magnetic field configuration that minimizes
total magnetic energy under the following requirements:
\begin{enumerate}
\item For $r>R_2$, the magnetic field is that of a dipole of dipole moment
  $\vec{m}$, inclined by $\vartheta_2$ with respect to the vertical axis
\item For $r<R_1$, the field is strictly vertical.
\end{enumerate}
Minimizing the magnetic energy in the transition layer $R_1<r<R_2$ implies
that $\mathbf{B}$ is a potential field in that region.

To find the minimum energy field satisfying these requirements, we
represent the magnetic field through scalar potentials $S$ and $T$:
\begin{equation}
  \vec{B} = -\nabla\times(\vec{x}\times\nabla S) -\vec{x}\times\nabla T \; .
\end{equation}
Both the vertical field and the inclined dipole field do not have a
toroidal part $T$, thus we set $T=0$ everywhere.
For the poloidal scalar potential $S$, we make the ansatz
\begin{equation} \label{Eq-S-spher-harm}
   S = \sum_{m=-1}^1 \left( a_1^m r + \frac{b_1^m}{r^2} \right)
                    Y_1^m(\vartheta,\varphi) \; .
        \left( \right) \; ,
\end{equation}
where $Y_1^m(\vartheta,\varphi)$ are the spherical harmonics of order $1$
and degree $m$.
Equation~(\ref{Eq-S-spher-harm}) is compatible with both the inner and the
outer field; any spherical harmonics of higher order would only increase
the total energy, so we do not include them.

Minimizing the sum of the energy (\ref{Eq-E-inner}) [with $R=R_1$] and the
energy of the potential field for $R_1<r<R_2$, we find the following
coefficients for $S$ after a straight-forward, but somewhat tedious
calculation:
\begin{equation}
  a_1^m = \tilde{a}_m \beta_1^m \; , \qquad
  b_1^m = \tilde{b}_m \beta_1^m \; ,
\end{equation}
where
\begin{equation}
  \beta_1^m
  \ =\ \frac{\mu_0|\vec{m}|}{4\pi} \sqrt{\frac{4\pi}{3}}
       \left\{\begin{array}{rl}
         \cos\vartheta_2    & \mbox{, $m=0$} \\[1.5ex]
         \mp\displaystyle\frac{\sin\vartheta_2}{\sqrt{2}}
                            & \mbox{, $m=\pm1$}
       \end{array}
       \right.  \; ,
\end{equation}
and
\begin{eqnarray}
  \tilde{a}_0
  &=&
  \left\{
    \begin{array}{ll}
      \dfrac{1}{R_2^3}  &, r<R_1\\
      \dfrac{1}{R_2^3}  &, R_1<r<R_2\\
      0\phantom{ABCDEi} &, r>R_2\\
    \end{array}
  \right. \\
  \tilde{a}_{\pm1}
  &=&
  \left\{
    \begin{array}{ll}
      0                      &, r<R_1\\
      \dfrac{1}{R_2^3-R_1^3} &, R_1<r<R_2\\
      0\phantom{ABCDEi}       &, r>R_2\\
    \end{array}
  \right. \\
  \tilde{b}_0
  &=&
  \left\{
    \begin{array}{ll}
      0               &, r<R_1\\
      0               &, R_1<r<R_2\\
      1\phantom{ABCDEi} &, r>R_2\\
    \end{array}
  \right. \\
  \tilde{b}_{\pm1}
  &=&
  \left\{\begin{array}{ll}
      0                           &, r<R_1\\
      -\dfrac{R_1^3}{R_2^3-R_1^3} &, R_1<r<R_2\\
      1                           &, r>R_2\\
    \end{array}
  \right.
\end{eqnarray}

To obtain the magnetic vector potential (which we need for the
\textsc{Pencil Code}), we use the formula
\begin{equation}
  \vec{A} = - \vec{x} \times\nabla S \; .
\end{equation}

For a thin transition layer of thickness
$\varepsilon \equiv R_2{-}R_1 \ll R_1$, the contributions to the magnetic
energy are
\begin{eqnarray}
  E(r>R_2)     &=& E_2 \equiv \dfrac{\mu_0 \vec{m}^2}{12\,\pi\,R_2^3} \; ,\\
  E(r_1<r<R_2) &=& \dfrac{\sin^2\vartheta_2}{\varepsilon} E_2 + O(1) \; , \\
  E(r>R_2)     &=& 2 \cos^2\vartheta_2 E_2 + O(\varepsilon) \; .
\end{eqnarray}


\begin{thebibliography}{}

\bibitem[Alcock, Farhi, \& Olinto 1986]{alcock86} Alcock, C., Farhi, E., \& Olinto, A. 1986,
 ApJ, 310, 261

\bibitem[Alford, Rajagopal, \& Wilczek(1998)]{alf:1998} Alford, M. G., Rajagopal, K. \& Wilczek, F., Phys. Lett. B 422 (1998), p. 247

\bibitem[Alford, Rajagopal, \& Wilczek(1999)]{alf:1999} Alford, M. G., Rajagopal, K. \& Wilczek, F., Nucl. Phys. B 537 (1999), p. 443

\bibitem[Alford, Berges, \& Rajagopal (2000)]{ABR2000} 
Alford, M., Berges, J., \& Rajagopal, K. 2000, Nucl. Phys. B, 571, 269 

\bibitem[Alford, Bowers, \& Rajagopal (2001)]{ABR2001} 
Alford, M., Bowers, J. A., \& Rajagopal, K. 2001, Phys. Rev. D, 63, 074016

\bibitem[Alpar (1991)]{alpar} Alpar, M. A. 1991, in Neutron Stars, Theory and Observation 
(Ventura, J. \& Pines, D.), Kluwer, Amsterdam, 1991, 49. 

\bibitem[Arnold, Moore, \& Yaffe (2000)]{AMY2000}
Arnold, P., Moore, G. D., \& Yaffe, L. G. 2000, Journal of High Energy 
Physics, 11, 001   

\bibitem[Bailin \& Love (1984)]{BL1984} 
Bailin, D., \& Love, A. 1984, Phys. Rep., 107, 325 

\bibitem[Baring \& Harding(2001)]{baring01}
 Baring, M. G., \& Harding, A. K. 2001, ApJ, 547, 929

\bibitem[Barrois(1977)]{bar:1977} Barrois, B. C. 1977, Nucl. Phys. B 129, 390.

\bibitem[Baym \& Heiselberg (1997)]{BH1997} 
Baym, G., \& Heiselberg, H. 1997, Phys. Rev. D, 56, 5254 

\bibitem[Blaschke, Grigorian, \& Voskresensky (2001)]{BGV2001}
Blaschke, D., Grigorian, H., \& Voskresensky, D. 2001, A\&A, 368, 561

\bibitem[Bombaci \& Datta(2000)]{2000ApJ...530L..69B} Bombaci, I., \& 
Datta, B.\ 2000, ApJ, 530, L69 

\bibitem[Carter \& Reddy (2000)]{CR2000} 
Carter, G. W., \& Reddy, S. 2000, Phys. Rev. D, 62, 103002 

\bibitem[Chatterjee, Hernquist, \& Narayan(2000)]{2000ApJ...534..373C} 
Chatterjee, P., Hernquist, L., \& Narayan, R.\ 2000, ApJ, 534, 373 

\bibitem[Chevalier(1989)]{Chev89} Chevalier, R. N. 1989, ApJ, 346, 847

\bibitem[Cline et al.(1980)]{1980ApJ...237L...1C} Cline, T.~L.~et al.\
1980, ApJ, 237, L1

\bibitem[Coe et al.(1994)]{Coe94} Coe, M. J., Jones, L. R., \& Lehto, H. 1994,
 MNRAS, 270, 178

\bibitem[{Collins \& Perry(1975)}]{col:1975} Collins, J. C. \& Perry, M. J., Phys. Rev. Lett. 34 (1975), p. 1353.
 
\bibitem[Corbett et al.(1995)]{Corbett95} Corbett, R. H. D., et al. 1995, ApJ, 443, 786

\bibitem[Dobler et al.(2006)]{dob:convstar}
Dobler, W., Stix, M. \& Brandenburg, A.\ 2006, ApJ, (accepted). 

\bibitem[Duncan \& Thompson(1992)]{1992ApJ...392L...9D} Duncan, R.~C.~\& Thompson, C.\ 1992, ApJ, 392, L9 

\bibitem[Fenimore, Klebesadel, \& Laros(1996)]{1996ApJ...460..964F}
Fenimore, E.~E., Klebesadel, R.~W., \& Laros, J.~G.\ 1996, ApJ, 460, 964

\bibitem[Feroci, Hurley, Duncan, \& Thompson(2001)]{2001ApJ...549.1021F}
Feroci, M., Hurley, K., Duncan, R.~C., \& Thompson, C.\ 2001, ApJ, 549,
1021

\bibitem[Fukue(2001)]{2001PASJ...53..687F} Fukue, J.\ 2001, PASJ, 53, 687 

\bibitem[Gaensler, Slane, Gotthelf, \& Vasisht(2001)]{2001ApJ...559..963G} 
Gaensler, B.~M., Slane, P.~O., Gotthelf, E.~V., \& Vasisht, G.\ 2001, ApJ, 
559, 963 

\bibitem[Gaensler (2002)]{G2002} Gaensler, B.~M.\ 2002, astro-ph/0212086 

\bibitem[Gaensler et al.(2005)]{2005ApJ...620L..95G} Gaensler, B.~M., 
McClure-Griffiths, N.~M., Oey, M.~S., Haverkorn, M., Dickey, J.~M., \& 
Green, A.~J.\ 2005, ApJ, 620, L95 

\bibitem[Gavriil, Kaspi, \& Woods(2002)]{2002Natur.419..142G} Gavriil, 
F.~P., Kaspi, V.~M., \& Woods, P.~M.\ 2002, Nature, 419, 142 

\bibitem[Glendenning(1997)]{glendenning97} Glendenning, N. K. 1997, Compact stars (Springer)

\bibitem[Goldreich \& Reisenegger(1992)]{Goldreich92} Goldreich, P. \&
 Reisenegger, A. 1992, ApJ, 395, 250

\bibitem[Golenetskij, Mazets, Ilinskij, \& Guryan(1979)]{1979SvAL....5..340G} 
Golenetskij, S.~V., Mazets, E.~P., Ilinskij, V.~N., \& Guryan, Y.~A.\ 1979, 
Soviet Astronomy Letters, 5, 340 

\bibitem[Gonzalez \& Safi-Harb(2003)]{Gonzalez03} Gonzalez, M., \& Safi-Harb,
 S. 2003, ApJ, 591, L143
 
 \bibitem[Gonzalez et al.(2004)]{Gonzalez04} Gonzalez, M. et al. 2004,
  ApJ, 605, 368

\bibitem[Graham-Smith(2003)]{graham03}
 Graham-Smith, F. 2003, Rep. Prog. Phys. 66, 173

\bibitem[Haensel (1991)]{H1991} 
Haensel, P. 1991, Nucl. Phys. B (Proc. Suppl.), 24, 23 

\bibitem[Heger, et al.(2004)]{heger} Heger, A., Woosley, S. E., \& Spruit, H. C. 2004, ApJ, submitted (astro-ph/0409422)
 
\bibitem[Heiselberg(1993)]{heiselberg93} Heiselberg, H., \& Pethick, C. J.\ 1993,
Phys. Rev. D, 48, 2916

\bibitem[Horvath, Benvenuto, \& Vucetich (1991)]{HBV1991} 
Horvath, J. E., Benvenuto, O. G., \& Vucetich, H. 1991, Phys. Rev. D,
44, 3797

\bibitem[Hurley et al.(1999a)]{1999Natur.397...41H} Hurley, K.~et al.\
1999a, Nature, 397, 41

\bibitem[Hurley et al.(1999b)]{1999ApJ...510L.111H} Hurley, K.~et al.\ 1999b, 
ApJ, 510, L111 

\bibitem[Ibrahim et al.(2001)]{2001ApJ...558..237I} Ibrahim, A.~I.~et al.\ 
2001, ApJ, 558, 237 

\bibitem[Iida \& Baym(2002)]{2002PhRvD..66a4015I} Iida, K.~\& Baym, G.\ 
2002, PRD, 66, 14015 

\bibitem[Iwamoto (1980)]{I1980} Iwamoto, N. 1980, Phys. Rev. Lett., 44, 1637 

\bibitem[Iwamoto (1982)]{I1982} Iwamoto, N. 1982, Ann. Phys., 141, 1 

\bibitem[Iwasawa, Koyama, \& Halpern(1992)]{Iwasawa92} Iwasawa, K., Koyama, K., \& Halpern, J. P. 1992, PASJ, 44, 9

\bibitem[Kaplan(2000)]{2000} Kaplan, D. L., 2000, in Spin, Magnetism and Cooling
 of Young Neutron Stars, http://online.itp.ucsb.edu/online/neustars$_{-}$c00/

\bibitem[Kaplan(2002)]{2002MmSAI..73..496K} Kaplan, D.~L.\ 2002, Memorie 
della Societa Astronomica Italiana, 73, 496 

\bibitem[Kaspi et al.(2003)]{kaspi03} Kaspi, V. M. et al., 2003, Ap.J., 588, L93. 

\bibitem[Kaspi(2004)]{kaspi04} Kaspi, V. M., 2004, astro-ph/0402175

\bibitem[Kaspi \& McLaughlin(2005)]{Kaspi05} Kaspi, V.  \&
 McLaughlin, M. A. 2005, ApJ, 618, L41

\bibitem[Ker\"anen, \& Ouyed(2003)]{keranen03} Ker\"anen, P., \& Ouyed, R. 2003, A\&A, 407, L51.  

\bibitem[Ker\"anen, Ouyed, \& Jaikumar(2005)]{keranen05} Ker\"anen, P., Ouyed, R., \& Jaikumar, P. 2005, ApJ, 618, 485

\bibitem[Kouveliotou et al.(1998)]{1998Natur.393..235K} Kouveliotou, C.~et 
al.\ 1998, Nature, 393, 235 

\bibitem[Kouveliotou et al.(1999)]{1999ApJ...510L.115K} Kouveliotou, C.~et 
al.\ 1999, ApJ, 510, L115 

\bibitem[Kriss et al.(1985)]{Kriss85} Kriss, G. A. et al. 1985, ApJ, 288, 703

\bibitem[Kulkarni et al.(2003)]{2003ApJ...585..948K} Kulkarni, S.~R., 
Kaplan, D.~L., Marshall, H.~L., Frail, D.~A., Murakami, T., \& Yonetoku, 
D.\ 2003, ApJ, 585, 948 

\bibitem[Lorimer et al.(1998)]{Lorimer98} Lorimer, D. R.,  Lyne, A. G., \& Camilo,
 F. 1998, A\&A, 331, 1002

\bibitem[Manchester \& Taylor(1977)]{1977QB843.P8M36....} Manchester, 
R.~N.~\& Taylor, J.~H.\ 1977, San Francisco : W.~H.~Freeman, c1977., 36 

\bibitem[Mazets et al.(1979)]{1979Natur.282..587M} Mazets, E.~P., 
Golenetskii, S.~V., Ilinskii, V.~N., Aptekar, R.~L., \& Guryan, I.~A.\ 1979, 
Nature, 282, 587 

\bibitem[Mazets et al.(1999)]{1999AstL...25..635M} Mazets, E.~P., Cline,
T.~L., Aptekar', R.~L., Butterworth, P.~S., Frederiks, D.~D., Golenetskii,
S.~V., Il'inskii, V.~N., \& Pal'shin, V.~D.\ 1999, Astronomy Letters, 25,
635

\bibitem[McLaughlin et al.(2003)]{McLaughlin03} McLaughlin, M. A. et al.
2003, ApJ, 591, L135

\bibitem[Meissner \& Ochsenfeld(1933)]{meissner} Meissner, W., Ochsenfeld, R., 1933, Naturwiss., 21, 787

\bibitem[Merghetti et al.(2000)]{Mereghetti00} Mereghetti, S. 2000, in The Neutron Star-Black Hole Connection, ed. V. Connaughton, C. Kouveliotou, J. Van Paradijs \& J. Ventura (NATO-ASI), in press (astro-ph/9911252)

\bibitem[Ouyed(2002)]{ouyed021} Ouyed, R., Dey, J., \& Dey, M. 2002, A\&A, 390, L39 

\bibitem[Ouyed et al.(2002)]{ouyed02} Ouyed, R., Elgar{\o}y, {\O}, Dahle, H., \& Ker\"anen, P. 2004, A\&A, 420, 1025

\bibitem[Paczynski(1992)]{paczynski92} Paczy\'nski, B. 1992, Acta Astron., 42, 145

\bibitem[Page \& Usov(2002)]{pu2002} 
Page, D., \& Usov, V.~V.\ 2002, PRL, 89, 1311011

\bibitem[Pisarski \& Rischke(1999)]{pis:1999} Pisarski, R. D. \& Rischke, D. H., Phys. Rev. Lett. 83 (1999), p. 37

\bibitem[Pivovaroff et al.(2000)]{Pivovaroff00} Pivovaroff, M., Kaspi, V.,
 \& Camilo, F. 2000, ApJ, 535, 379

\bibitem[Rapp et al.(1998)]{rapp:1998} Rapp, R., Sch\"afer, T., Shuryak, E. V., \& Velkovsky, M., Phys. Rev. Lett. 81 (1998), p. 53

\bibitem[Rischke(2000a)]{rischke00a} Rischke, D. H., 2000, Phys. Rev. D 62, 034007

\bibitem[Rischke(2000b)]{rischke00b}  Rischke, D. H., 2000, Phys. Rev. D 62, 054017

\bibitem[Ruutu et al.(1997)]{1997PhRvB..5614089R} Ruutu, V.~M., Ruohio, 
J.~J., Krusius, M., Pla{\c c}ais, B., Sonin, E.~B., \& Xu, W.\ 1997, PRB, 
56, 14089 

\bibitem[Rybicki \& Lightman(1979)]{Rybicki79} Rybicki, G. B., \& Lightman,
A. P. in Radiative Processes in Astrophysics, 1979 (Wiley \& Sons).

\bibitem[Sch\"afer(2003)]{sch:2003} Sch\"afer, T., (hep-ph/0304281).

\bibitem[Sedrakian, Blaschke, Shahabasyan, \& Voskresensky]{sbsv2001} 
Sedrakian, D. M., Blaschke, D., Shahabasyan, K. M., \& Voskresensky, 
D. N. 2001, Astrofizika, 44, 443 

\bibitem[Shovkovy \& Ellis (2003)]{SE2003}
Shovkovy, I. A., \& Ellis, P. J. 2003, Phys. Rev. C, 67, 048801

\bibitem[Son(1999)]{son:1999} Son, D. T., Phys. Rev. D 59 (1999), p. 094019

\bibitem[Thompson \& Duncan(1995)]{1995MNRAS.275..255T} Thompson, C.~\& 
Duncan, R.~C.\ 1995, MNRAS, 275, 255 

\bibitem[Thompson \& Duncan(1996)]{1996ApJ...473..322T} Thompson, C.~\& 
Duncan, R.~C.\ 1996, ApJ, 473, 322 

\bibitem[Tilley \& Tilley(1990)]{TT90} Tilley, D. R., \& Tilley, J. 
Superfluidity and Superconductivity, 3rd ed., Bristol, England, Hilger 1990

\bibitem[Tsubota, Kasamatsu, \& Ueda (2002)]{TKU2002}
Tsubota, M., Kasamatsu, K., \& Ueda, M. 2002, Phys. Rev. D, 65, 023603


\bibitem[Usov(2001)]{2001PhRvL..87b1101U} Usov, V.~V.\ 2001, Physical Review Letters, 87, 021101 

\bibitem[van Kerkwijk \& Kulkarni(2001)]{2001A&A...380..221V} van Kerkwijk, 
M.~H., \& Kulkarni, S.~R.\ 2001, A\&A, 380, 221 

\bibitem[Vancura, Blair, Long, \& Raymond(1992)]{1992ApJ...394..158V} 
Vancura, O., Blair, W.~P., Long, K.~S., \& Raymond, J.~C.\ 1992, ApJ, 394, 
158 

\bibitem[Woods et al.(1999)]{1999ApJ...527L..47W} Woods, P.~M.~et al.\ 
1999, ApJ, 527, L47 

\bibitem[Woods et al.(2002)]{2002ApJ...576..381W} Woods, P.~M., 
Kouveliotou, C., G{\" o}{\u g}{\" u}{\c s}, E., Finger, M.~H., Swank, J., 
Markwardt, C.~B., Hurley, K., \& van der Klis, M.\ 2002, ApJ, 576, 381 

\bibitem[Woods (2003)]{Woods03a} Woods, P. M., 2003, astro-ph/0304372 

\bibitem[Woods et al. (2003)]{Woods03b} Woods, P. M. et al., 2003, 
astro-ph/0310575

\bibitem[Zane et al.(2001)]{Zane01} Zane, S., Turolla, R., Stella, L., \& Treves, A.
 2001, ApJ, 560, 384

\bibitem[Zhang \& Harding(2000)]{Zhang00} Zhang, B., \& Harding, A. K. 2000,
 ApJ, 535, L51

\end{thebibliography}
\end{document}